# Robust Input Shaping Control for Flexible Structures Based on Unscented Kalman Filter

*Weiyi Yang[1], Yu Yuan[2, *], Mingsheng Shang[3]*

[1]Chongqing Institute of Green and Intelligent Technology, Chinese Academy of Sciences, Chongqing 400714, China, and also Chongqing School, University of Chinese Academy of Sciences, Chongqing 400714, China
[2] College of Computer and Information Science, Southwest University, Chongqing 400715, China
[3] Chongqing Institute of Green and Intelligent Technology, Chongqing Key Laboratory of Edge AI Computing, Chinese Academy of Sciences, Chongqing 400714, China
*yuanyekl@gmail.com

**Abstract**

With the rapid development of industrial automation and smart manufacturing, the control of flexible structures and underactuated systems has become a critical research focus. Residual vibrations in these systems not only degrade operational efficiency but also pose risks to structural integrity and longevity. Traditional input shaping techniques, while effective, often suffer from performance degradation due to parameter inaccuracies and environmental disturbances. To address these challenges, this paper introduces an innovative <u>u</u>nscented Kalman filter-based <u>z</u>ero vibration derivative input <u>s</u>haping (UZS) method. The proposed approach combines two key innovations: 1) a data-driven Unscented Kalman Filterfor real-time system parameter identification, and 2) a zero-vibration derivative (ZVD) input shaper for robust vibration suppression. To validate the effectiveness of UZS, we conducted extensive experiments on a vertical flexible beam platform, and the results demonstrate significant improvements over state-of-the-art methods. Additionally, we have made the experimental datasets publicly available to facilitate further research. The findings highlight UZS's potential for practical applications in industrial automation, robotics, and precision engineering.

## 1   Introduction

The accelerating development of industrial automation and smart manufacturing has established dynamic control of flexible underactuated systems as a pivotal research domain [1-4]. Underactuated mechanical systems, particularly cranes [5-8], have drawn substantial research interest due to their critical role in port operations, construction, and logistics [9-14]. A fundamental challenge in these systems involves vibration suppression during high-speed operation, where excessive oscillations impair positioning accuracy, reduce operational efficiency, and potentially compromise structural integrity [15-18]. Consequently, developing effective vibration control methods carries both theoretical significance and practical engineering value [19-24].

Traditional vibration suppression approaches primarily rely on structural optimization or classical control theory [25-27]. However, advancements in sensor technology and data acquisition have revealed the considerable potential of data-driven control methods [28-33]. Modern sensing systems now provide higher measurement precision at reduced costs, enabling more accurate state estimation and system modeling [34-38]. This technological progress is driving a shift from model-based designs toward intelligent, data-driven control strategies for vibration mitigation [39-41].

Various feedback control techniques, i.e., including adaptive control [42], sliding mode control [43], fuzzy logic control [44], and model predictive control (MPC) [45], have demonstrated notable success in enhancing the stability and robustness of flexible systems. Despite their effectiveness, practical deployment remains constrained by real-world challenges. Industrial controllers are often closed, proprietary platforms with limited configurability [46], and the reliance on high-precision sensors introduces additional complexity and cost, hindering the adoption of advanced feedback control strategies in industrial environments [47]. These limitations have spurred growing interest in feedforward control techniques such as input shaping (IS) [48], signal filtering [49], and trajectory smoothing [50], which offer simpler implementation. IS has emerged as particularly effective for vibration suppression in flexible systems due to its model-independent nature and adaptability to operational conditions [51-55]. However, control performance in flexible nonlinear systems remains highly sensitive to parameter accuracy, with uncertainties and drifts frequently arising from structural variations or external disturbances [56-58]. To improve robustness and maintain control performance, researchers have proposed a variety of parameter identification and optimization techniques [59-61]. Early work largely focused on optimization-based methods. For example, Yi et al. [62] applied genetic algorithms (GA) to optimize system responses, successfully mitigating the impact of parameter uncertainty. However, the complexity of such algorithms frequently results in slow convergence. In contrast, particle swarm optimization (PSO) [63] has gained popularity for its superior convergence speed, especially when integrated with conventional control strategies [64]. Yet, PSO is prone to premature convergence to local optima, particularly under improper parameter tuning or repeated iterations, which can degrade its parameter estimation accuracy [65-68]. To overcome these drawbacks, Li et al. [69]

introduced a hybrid method that incorporates recursive least squares to confine the PSO search space, thereby improving both efficiency and estimation accuracy.

With the growing integration of artificial neural networks (ANNs) in control and system modeling, PSO has also been used to generate high-quality training datasets for neural networks, boosting model generalization and prediction accuracy [70-76]. Nevertheless, ANN-based approaches typically require large datasets and may exhibit poor performance in data-scarce scenarios [77-85]. Motivated by these insights, this study proposes a novel <u>u</u>nscented Kalman filter-based <u>z</u>ero vibration derivative input <u>s</u>haping (UZS) method. The proposed approach employs the unscented Kalman filter (UKF) to dynamically estimate and correct parameter deviations, thereby enhancing the robustness and effectiveness of the input shaper and enabling precise vibration suppression under real-world conditions. To support reproducibility and facilitate further research, a comprehensive dataset based on a vertical flexible beam system has been constructed and made publicly available.

Experimental results demonstrate that the UZS method significantly outperforms several state-of-the-art input shaping techniques in suppressing vibrations, underscoring its strong potential for real-world deployment in industrial control systems.

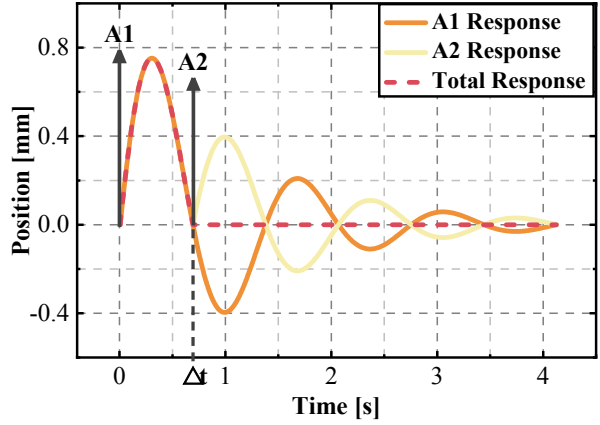

Fig.1 The principle of IS. The residual vibration generated by A1 is effectively suppressed by A2 and has an appropriate amplitude and delay.

## 2 PRELIMINARIES

*2.1 Theoretical foundation of IS*

IS represents a classical feedforward open-loop control strategy designed to mitigate system vibrations through convolution of the original input signal with a specifically designed impulse sequence [76-79]. As illustrated in Fig 1, this methodology proves particularly effective for undamped or underdamped second-order systems, characterized by the standard transfer function as follows:

$$G(s) = \frac{\omega_n^2}{s^2 + 2\zeta\omega_n s + \omega_n^2}, \quad (1)$$

where $\omega_n$ denotes the undamped natural frequency and $\zeta$ represents the damping ratio. In such systems, a unit impulse input at time $t_n$ generates a predictable dynamic response. The IS technique operates by convolving the original command signal with multiple time-delayed impulses, transforming the system response into a superposition of individual impulse responses. Through trigonometric transformation, this composite response can be expressed as:

$$y(t) = \frac{\omega_n}{\sqrt{1-\zeta^2}} e^{-\zeta\omega_n t} \sqrt{C^2(\zeta,\omega_n) + S^2(\zeta,\omega_n)} \sin(\omega_d t - \varphi), \quad (2)$$

where $A_i$ and $t_i$ correspond to the amplitude and temporal position of the $i$-th impulse respectively, $N$ indicates the total number of impulses, and $\varphi$, $C(\zeta, \omega_n)$ and $S(\zeta, \omega_n)$ are defined as:

$$\begin{cases} C(\zeta,\omega_n) = \sum_{i=1}^{n} A_i e^{\zeta\omega_n t_i} \cos\omega_d t_i \\ S(\zeta,\omega_n) = \sum_{i=1}^{n} A_i e^{\zeta\omega_n t_i} \sin\omega_d t_i \end{cases}, \quad \varphi = \arctan\frac{C(\zeta,\omega_n)}{S(\zeta,\omega_n)}, \quad (3)$$

To ensure both desired motion execution and residual vibration suppression, the following normalization constraint must be satisfied:

$$\sum_{i=1}^{N} A_i = 1. \quad (4)$$

The magnitude of residual vibration can be quantified by the residual vibration ratio, i.e., the ratio between the final system response and the ideal (vibration-free) response, which is defined as:

$$V(\zeta, \omega_n) = e^{-\zeta \omega_n t_N} \sqrt{C^2(\zeta, \omega_n) + S^2(\zeta, \omega_n)}. \quad (5)$$

From a theoretical standpoint, perfect vibration cancellation can be achieved when the input shaper is constructed such that the resulting convolution with the system's impulse response leads to complete destructive interference of all residual vibration modes. Mathematically, this requires the shaping impulses to be designed such that the weighted sum of modal responses satisfies both amplitude and phase conditions that eliminate oscillatory components. Accordingly, the fundamental task in input shaper design is the determination of an optimal set of amplitude and time pairs [$A_i$, $t_i$] that collectively minimize residual vibration while preserving system responsiveness.

To enhance robustness in the presence of modeling inaccuracies, parameter drift, and external disturbances, a variety of advanced shaping techniques have been proposed. Notable among these are the Zero Vibration (ZV) shaper, which cancels vibration under nominal conditions; the Zero Vibration Derivative-Derivative (ZVDD) shaper, which additionally attenuates sensitivity to natural frequency variations; and the Enhanced Insensitivity (EI) shaper, designed to maintain performance over a broader range of system uncertainties [86-95]. These methods represent critical developments in input shaping theory, particularly for underdamped and flexible dynamic systems where precise vibration mitigation is essential. In this study, we adopt the ZVD shaper to suppress residual vibrations, and its impulse sequence is defined as:

$$\text{ZVD} = \begin{bmatrix} A_i \\ t_i \end{bmatrix} = \begin{bmatrix} \dfrac{1}{(1+K)^2} & \dfrac{2K}{(1+K)^2} & \dfrac{K^2}{(1+K)^2} \\ 0 & \dfrac{\pi}{\omega_d} & \dfrac{2\pi}{\omega_d} \end{bmatrix}. \quad (6)$$

Although various input shaping techniques can theoretically achieve complete suppression of residual vibrations under ideal parameter conditions, their practical performance is often substantially compromised by structural variations, external disturbances, and measurement noise. These factors introduce parameter uncertainties that degrade the effectiveness of the shaper in real-world scenarios. To address this limitation and enhance the robustness of vibration control under uncertain conditions, the following section introduces the UKF method. This approach facilitates data-driven estimation of key system parameters and compensates for their deviations in real time, thereby improving the stability, accuracy, and reliability of the input shaper in dynamic and uncertain environments.

*2.1 The UKF algorithm*

The Unscented Kalman Filter (UKF) employs the Unscented Transformation (UT) to propagate the mean and covariance of the state distribution through nonlinear system dynamics, thereby avoiding the approximation and linearization errors commonly associated with the Extended Kalman Filter (EKF) [76].

At each filtering step, $2n+1$ sigma points are generated based on the current state estimate $\mathbf{x}_k \in R^n$ and its covariance $\mathbf{P}_k \in R^{n \times n}$, which is given as:

$$\begin{cases} \chi_k^{(0)} = \mathbf{x}_k, \\ \chi_k^{(i)} = \mathbf{x}_k + \sqrt{(n+\lambda)\mathbf{P}_{ki}}, & i=1,...,n \\ \chi_k^{(i+1)} = \mathbf{x}_k - \sqrt{(n+\lambda)\mathbf{P}_{ki}}, \end{cases} \quad (7)$$

where $\lambda = \alpha^2(n+\kappa) - n$ is a scaling factor, and $\alpha$, $\kappa$ are user-defined parameters. The square root is computed via Cholesky decomposition. Each sigma point is propagated through the nonlinear state transition function $\Gamma(\cdot)$, which is expressed as:

$$\chi_{k+1|k}^{(i)} = \Gamma\left(\chi_k^{(i)}\right). \quad (8)$$

The predicted state mean and covariance are computed as:

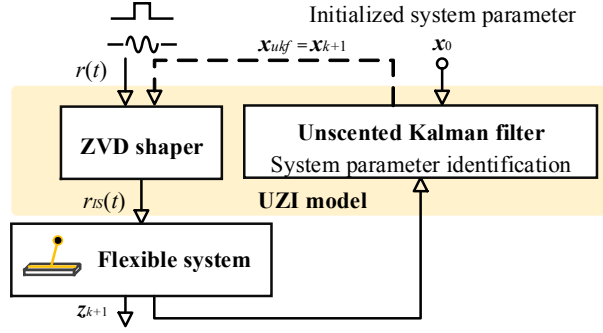

Fig. 2 The workflow of the proposed UZS method.

$$\hat{\boldsymbol{x}}_{k+1|k} = \sum_{i=0}^{2n} w_i^{(m)} \chi_{k+1|k}^{(i)},$$
$$\mathbf{P}_{k+1|k} = \sum_{i=0}^{2n} w_i^{(c)} \left( \chi_{k+1|k}^{(i)} - \hat{\boldsymbol{x}}_{k+1|k} \right)\left( \chi_{k+1|k}^{(i)} - \hat{\boldsymbol{x}}_{k+1|k} \right)^T + \mathbf{Q}, \quad (9)$$

where $\mathbf{Q}$ is the process noise covariance, and $w_i^{(m)}$, $w_i^{(c)}$ are weights for the mean and covariance. The predicted sigma points are then passed through the observation function $\boldsymbol{H}(\cdot)$:

$$\mathbf{z}_{k+1|k}^{(i)} = \boldsymbol{H}\left( \chi_{k+1|k}^{(i)} \right). \quad (10)$$

From these, the predicted observation mean and covariance can be calculated as:

$$\hat{\mathbf{z}}_{k+1|k} = \sum_{i=0}^{2n} w_i^{(m)} \mathbf{z}_{k+1|k}^{(i)},$$
$$\mathbf{S}_{k+1} = \sum_{i=0}^{2n} w_i^{(c)} \left( \mathbf{z}_{k+1|k}^{(i)} - \hat{\mathbf{z}}_{k+1|k} \right)\left( \mathbf{z}_{k+1|k}^{(i)} - \hat{\mathbf{z}}_{k+1|k} \right)^T + \mathbf{R}, \quad (11)$$

where $\mathbf{R}$ is the observation noise covariance, and the cross-covariance between the state and observation is computed as:

$$\mathbf{P}_{xz} = \sum_{i=0}^{2n} w_i^{(c)} \left( \chi_{k+1|k}^{(i)} - \hat{\boldsymbol{x}}_{k+1|k} \right)\left( \mathbf{z}_{k+1|k}^{(i)} - \hat{\mathbf{z}}_{k+1|k} \right)^T. \quad (12)$$

This is used to calculate the Kalman gain:

$$\mathbf{K}_{k+1} = \mathbf{P}_{xz} \mathbf{S}_{k+1}^{-1}. \quad (13)$$

Based on the actual measurement $\mathbf{z}_{k+1}$, the state estimate and covariance are updated as:

$$\boldsymbol{x}_{k+1} = \hat{\boldsymbol{x}}_{k+1|k} + \mathbf{K}_{k+1}\left( \mathbf{z}_{k+1} - \hat{\mathbf{z}}_{k+1|k} \right),$$
$$\mathbf{P}_{k+1} = \mathbf{P}_{k+1|k} - \mathbf{K}_{k+1} \mathbf{S}_{k+1} \mathbf{K}_{k+1}^T. \quad (14)$$

To monitor convergence, the error function is defined as:

$$e = \sum_{k=1}^{|M|} \left\| \mathbf{z}_{k+1} - \mathbf{z}_{k+1|k}^{(i)} \right\|, \quad (15)$$

where $M$ is the actual vibration training set measured by a vibration displacement sensor. If the error $e$ falls below a predefined threshold, the iteration terminates. Otherwise, the algorithm proceeds to the next cycle. This method enables dynamic parameter tracking and enhances the robustness and accuracy of input shaping in the presence of model uncertainties. Based on the above rationale, the operational workflow of the proposed UZS method is depicted in Fig 2.

## 3  Experiment and Comparison

*3.1 General Settings*

*3.1.1 Evaluation Protocol:* We employ the following three metrics i.e., maximum error (MAX), root mean square error (RMSE) and mean error (MEAN) [30], to evaluate the performance of our method:

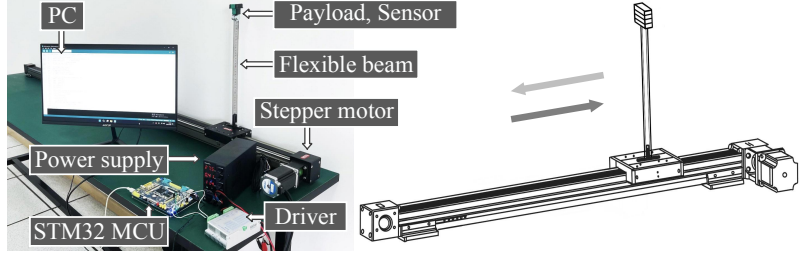

Fig. 3 The VFB experimental platform.

$$\text{MAX} = \max |r_i - \hat{r}_i|,$$
$$\text{RMSE} = \sqrt{\frac{1}{n}\sum_{i=1}^{n}(r_i - \hat{r}_i)^2}, \quad (16)$$
$$\text{MEAN} = \frac{1}{n}\sum_{i=1}^{n}|r_i - \hat{r}_i|.$$

*3.1.2 Datasets:* This study utilizes two experimental datasets (D1 and D2) collected from our laboratory-scale vertical flexible beam (VFB) platform, as illustrated in Figure 3. Two distinct VFB configurations are implemented to evaluate system performance under different dynamic conditions: VFB-1 with a 0.125 kg payload and 0.35 m beam length, and VFB-2 with a 0.09 kg payload and 0.45 m beam length. Each dataset contains comprehensive vibration characteristics including displacement responses, identified natural frequencies, and estimated damping ratios. All experimental data have been made publicly available on GitHub[1] to support research reproducibility. For robust performance evaluation, we implement a rigorous experimental protocol: (1) 400 samples were randomly selected from each dataset following a uniform distribution, with this sampling process repeated 10 times to minimize selection bias; (2) each sample set was partitioned into 90% training data and 10% testing data. Model performance is assessed based on the mean and standard deviation across all 10 trials. The training procedure incorporated two termination criteria: (i) maximum iteration count of 100 epochs, or (ii) improvement in the objective function below $10^{-4}$ between consecutive iterations, ensuring both computational efficiency and convergence reliability.

*3.1.3 Experimental Setup:* The vertical flexible beam (VFB) experimental platform consists of a stepper motor-driven slider assembly mounted on precision linear guide rails, capable of supporting up to 30 kg payload over a 2-meter travel range. vibration characteristics are precisely measured using a WTVB01-BT50 non-contact displacement sensor, while system control is implemented through an STM32H743 microcontroller (Arm Cortex-M7 core) programmed via Arduino IDE in a Windows 11 environment, ensuring reliable real-time control and data acquisition at configurable sampling rates up to 1 kHz.

*3.2 Performance Comparison*

*3.2.1 Compared Models:*
- M1: A single ZVD shaper [47].
- M2: An Levenberg-Marquard [48] algorithm optimized method.
- M3: A PSO [69] algorithm optimized method.
- M4: An ANN [70] optimized method, whose network structure is 2-20-20-6-1.
- M5: An EKF [76] optimized method.
- M6: The UZS method presented in the article.

*3.2.2 Compared results:* The experimental results reveal several key findings regarding the performance of the evaluated methods. As demonstrated across both D1 and D2 datasets, method M6 consistently outperforms all competing approaches (M1-M5) in terms of parameter identification efficiency. Statistical analysis using the Wilcoxon signed-rank test confirms this superiority with compelling evidence: M6 achieves the maximum sum of positive ranks ($R^+ = 78$) in all pairwise comparisons, supported by a highly significant p-value of 0.00024. This strong statistical evidence ($p < 0.001$) conclusively establishes M6's superior performance. The training dynamics of methods M2-M6 are visualized in Fig 9, which illustrates their convergence behavior on both datasets. More importantly, Fig 5 presents practical validation of the UZS method's control capabilities in a real-world motion task. The results clearly demonstrate UZS's exceptional vibration suppression performance during operation, confirming its advantages in both dynamic control precision and system robustness. These experimental outcomes collectively validate the effectiveness of the proposed approach across multiple performance metrics.

---

[1] https://github.com/YWYucas/U-Datasets

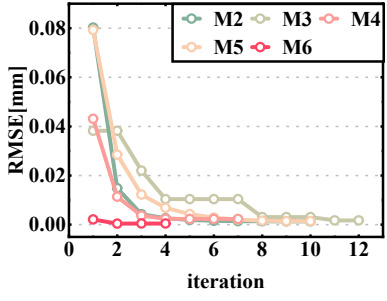
(a) RMSE on VFB-1

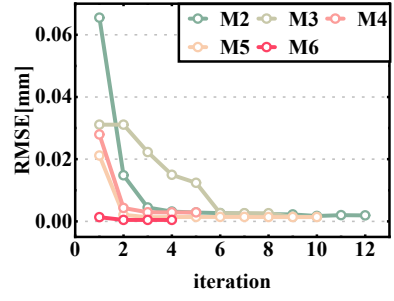
(b) RMSE on VFB-2

Fig. 4 Comparison of training curves of M2-M6 on D1-D2 in RMAE.

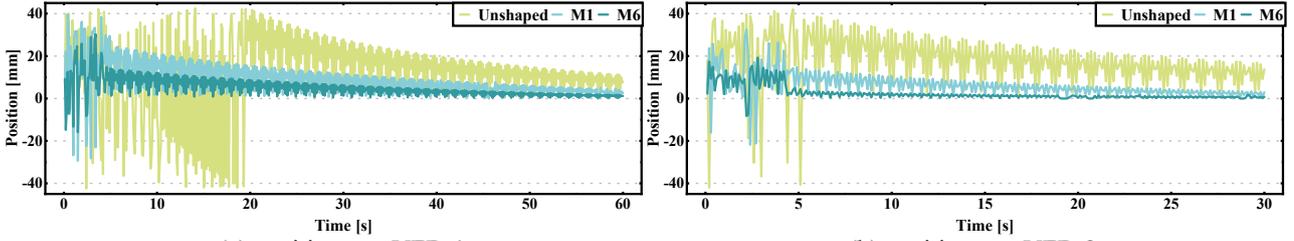

(a) positions on VFB-1    (b) positions on VFB-2

Fig. 5 Comparison of actual vibration positions on VFB.

Table 1 The results of the comparative experimental errors (in mm)

| Dataset | Metric | M1 | M2 | M3 | M4 | M5 | M6 |
|---|---|---|---|---|---|---|---|
| D1 | MAX | 40.00±0.70 | 4.91±2.38 | 4.52±1.11 | 4.18±1.72 | 4.16±0.77 | 3.11±0.84* |
| | RMSE | 37.77±0.15 | 6.20±1.76 | 3.27±1.28 | 1.38±0.23 | 1.64±0.36 | 1.30±0.34* |
| | MEAN | 37.76±0.15 | 1.37±0.63 | 1.58±0.53 | 1.16±0.14 | 1.33±0.40 | 1.02±0.33* |
| D2 | MAX | 32.53±0.11 | 5.10±1.86 | 4.50±1.04 | 4.41±0.52 | 4.81±1.09 | 3.01±0.67* |
| | RMSE | 30.75±0.22 | 5.99±1.43 | 2.70±0.93 | 1.90±0.12 | 1.88±0.63 | 1.50±0.18* |
| | MEAN | 30.71±0.23 | 1.89±0.81 | 1.23±0.25 | 1.31±0.10 | 1.49±0.59 | 1.19±0.18* |

## 4 Conclusion

To advance vibration control in industrial applications, this study introduces a novel UZS method that significantly enhances the practical implementation of input shaping techniques. The key innovation lies in integrating UKF for real-time system parameter identification, which effectively addresses the implementation challenges of conventional input shapers in practical settings. Extensive experimental validation on a flexible beam platform demonstrates that the proposed UZS method achieves superior vibration suppression performance compared to existing input shaping approaches, while maintaining computational efficiency. For future research directions, we identify two promising avenues: (1) development of accelerated optimization algorithms to enable real-time parameter adaptation in dynamic environments, and (2) design of robust input shaper architectures with enhanced disturbance rejection capabilities. These improvements aim to address current limitations in handling significant parameter variations and external disturbances, potentially expanding the applicability of input shaping techniques to more challenging industrial scenarios with stringent performance requirements